\documentclass[aps,prd,twocolumn,superscriptaddress,showpacs,preprintnumbers,amsmath,amssymb]{revtex4}
\usepackage{graphicx} 
\usepackage{dcolumn}  
\usepackage[]{lineno}

\usepackage{xcolor}

\graphicspath{{ps}}

\begin{document}




\title{ \quad\\[1.0cm] Measurement of the Decays $B^0_s\to
  J/\psi\,\phi(1020)$, $B^0_{s}\to J/\psi\,f'_2(1525)$\\ and $B^0_s\to J/\psi\,K^+K^-$ at Belle}

\noaffiliation
\affiliation{University of the Basque Country UPV/EHU, 48080 Bilbao}
\affiliation{Beihang University, Beijing 100191}
\affiliation{University of Bonn, 53115 Bonn}
\affiliation{Budker Institute of Nuclear Physics SB RAS and Novosibirsk State University, Novosibirsk 630090}
\affiliation{Faculty of Mathematics and Physics, Charles University, 121 16 Prague}
\affiliation{University of Cincinnati, Cincinnati, Ohio 45221}
\affiliation{Deutsches Elektronen--Synchrotron, 22607 Hamburg}
\affiliation{Department of Physics, Fu Jen Catholic University, Taipei 24205}
\affiliation{Justus-Liebig-Universit\"at Gie\ss{}en, 35392 Gie\ss{}en}
\affiliation{Gyeongsang National University, Chinju 660-701}
\affiliation{Hanyang University, Seoul 133-791}
\affiliation{University of Hawaii, Honolulu, Hawaii 96822}
\affiliation{High Energy Accelerator Research Organization (KEK), Tsukuba 305-0801}
\affiliation{Hiroshima Institute of Technology, Hiroshima 731-5193}
\affiliation{Ikerbasque, 48011 Bilbao}
\affiliation{Indian Institute of Technology Guwahati, Assam 781039}
\affiliation{Indian Institute of Technology Madras, Chennai 600036}
\affiliation{Institute of High Energy Physics, Chinese Academy of Sciences, Beijing 100049}
\affiliation{Institute of High Energy Physics, Vienna 1050}
\affiliation{Institute for High Energy Physics, Protvino 142281}
\affiliation{INFN - Sezione di Torino, 10125 Torino}
\affiliation{Institute for Theoretical and Experimental Physics, Moscow 117218}
\affiliation{J. Stefan Institute, 1000 Ljubljana}
\affiliation{Kanagawa University, Yokohama 221-8686}
\affiliation{Institut f\"ur Experimentelle Kernphysik, Karlsruher Institut f\"ur Technologie, 76131 Karlsruhe}
\affiliation{Korea Institute of Science and Technology Information, Daejeon 305-806}
\affiliation{Korea University, Seoul 136-713}
\affiliation{Kyungpook National University, Daegu 702-701}
\affiliation{\'Ecole Polytechnique F\'ed\'erale de Lausanne (EPFL), Lausanne 1015}
\affiliation{Faculty of Mathematics and Physics, University of Ljubljana, 1000 Ljubljana}
\affiliation{University of Maribor, 2000 Maribor}
\affiliation{Max-Planck-Institut f\"ur Physik, 80805 M\"unchen}
\affiliation{School of Physics, University of Melbourne, Victoria 3010}
\affiliation{Moscow Physical Engineering Institute, Moscow 115409}
\affiliation{Graduate School of Science, Nagoya University, Nagoya 464-8602}
\affiliation{Kobayashi-Maskawa Institute, Nagoya University, Nagoya 464-8602}
\affiliation{Nara Women's University, Nara 630-8506}
\affiliation{National Central University, Chung-li 32054}
\affiliation{National United University, Miao Li 36003}
\affiliation{Department of Physics, National Taiwan University, Taipei 10617}
\affiliation{H. Niewodniczanski Institute of Nuclear Physics, Krakow 31-342}
\affiliation{Nippon Dental University, Niigata 951-8580}
\affiliation{Niigata University, Niigata 950-2181}
\affiliation{University of Nova Gorica, 5000 Nova Gorica}
\affiliation{Osaka City University, Osaka 558-8585}
\affiliation{Pacific Northwest National Laboratory, Richland, Washington 99352}
\affiliation{Panjab University, Chandigarh 160014}
\affiliation{University of Pittsburgh, Pittsburgh, Pennsylvania 15260}
\affiliation{University of Science and Technology of China, Hefei 230026}
\affiliation{Seoul National University, Seoul 151-742}
\affiliation{Soongsil University, Seoul 156-743}
\affiliation{Sungkyunkwan University, Suwon 440-746}
\affiliation{School of Physics, University of Sydney, NSW 2006}
\affiliation{Tata Institute of Fundamental Research, Mumbai 400005}
\affiliation{Excellence Cluster Universe, Technische Universit\"at M\"unchen, 85748 Garching}
\affiliation{Tohoku Gakuin University, Tagajo 985-8537}
\affiliation{Tohoku University, Sendai 980-8578}
\affiliation{Department of Physics, University of Tokyo, Tokyo 113-0033}
\affiliation{Tokyo Institute of Technology, Tokyo 152-8550}
\affiliation{Tokyo University of Agriculture and Technology, Tokyo 184-8588}
\affiliation{University of Torino, 10124 Torino}
\affiliation{CNP, Virginia Polytechnic Institute and State University, Blacksburg, Virginia 24061}
\affiliation{Wayne State University, Detroit, Michigan 48202}
\affiliation{Yamagata University, Yamagata 990-8560}
\affiliation{Yonsei University, Seoul 120-749}
  \author{F.~Thorne}\affiliation{Institute of High Energy Physics, Vienna 1050} 
  \author{C.~Schwanda}\affiliation{Institute of High Energy Physics, Vienna 1050} 
  \author{I.~Adachi}\affiliation{High Energy Accelerator Research Organization (KEK), Tsukuba 305-0801} 
  \author{H.~Aihara}\affiliation{Department of Physics, University of Tokyo, Tokyo 113-0033} 
  \author{D.~M.~Asner}\affiliation{Pacific Northwest National Laboratory, Richland, Washington 99352} 
  \author{V.~Aulchenko}\affiliation{Budker Institute of Nuclear Physics SB RAS and Novosibirsk State University, Novosibirsk 630090} 
  \author{T.~Aushev}\affiliation{Institute for Theoretical and Experimental Physics, Moscow 117218} 
  \author{A.~M.~Bakich}\affiliation{School of Physics, University of Sydney, NSW 2006} 
  \author{A.~Bala}\affiliation{Panjab University, Chandigarh 160014} 
  \author{B.~Bhuyan}\affiliation{Indian Institute of Technology Guwahati, Assam 781039} 
% ...
\collaboration{The Belle Collaboration}

\begin{abstract}
We report a measurement of the branching fraction of the
  decay $B^0_s\to J/\psi\,\phi(1020)$, evidence and a branching fraction
  measurement for $B^0_s\to J/\psi\,f'_2(1525)$, and the
  determination of the total $B^0_s\to J/\psi\,K^+K^-$ branching fraction,
  including the resonant and non-resonant
  contributions to the $K^+K^-$~channel. We also determine the $S$-wave contribution within the 
$\phi(1020)$ mass region. The absolute branching fractions are
$\mathcal{B}[B^0_s\to J/\psi\,\phi(1020)]=(1.25 \pm 0.07\left(\mathrm{stat}\right)\pm
0.08\left(\mathrm{syst}\right)\pm 0.22\left(f_s\right))\times 10^{-3}$,
$\mathcal{B}[B^0_s\to J/\psi\,f'_2(1525)]=(0.26\pm 0.06\left(\mathrm{stat}\right)
\pm 0.02\left(\mathrm{syst}\right) \pm 0.05\left(f_s\right))\times 10^{-3}$ and
$\mathcal{B}[B^0_s\to J/\psi\,K^+K^-] = (1.01\pm
  0.09\left(\mathrm{stat}\right) \pm 0.10\left(\mathrm{syst}\right)\pm 0.18\left(f_s\right))\times
  10^{-3}$, where the last systematic error is due to the branching fraction of $b\bar{b}\to B^{(*)}_s B^{(*)}_s$. The branching fraction ratio 
is found to be $\mathcal{B}[B^0_s\to J/\psi\,f'_2(1525)]/\mathcal{B}[B^0_s\to
J/\psi\,\phi(1020)]=(21.5\pm 4.9\left(\mathrm{stat}\right)
\pm2.6\left(\mathrm{syst}\right))\%$. All results are based on a 121.4~fb$^{-1}$
data sample collected at the $\Upsilon(5S)$~resonance by the Belle
experiment at the KEKB asymmetric-energy $e^+e^-$ collider.\\

\end{abstract}


\maketitle

\tighten

{\renewcommand{\thefootnote}{\fnsymbol{footnote}}}
\setcounter{footnote}{0}

\section{Introduction}

The study of $B^0_s\bar B^0_s$ mixing and $CP$~violation in
$B^0_s$~decays~\cite{Dunietz:2000cr} helps advance our 
understanding of the Cabibbo-Kobayashi-Maskawa
mechanism~\cite{KM,Cab}. The decay $B^0_s\to
J/\psi\,\phi(1020)$~\cite{CC} probes the $CP$-violating phase
$\phi_s$ of $B^0_s\bar B^0_s$
oscillations~\cite{cdfswave,Abazov:2011ry,LHCb:2011aa,atlas_phi}, which is
predicted to be small within the Standard Model (SM). However,
contributions from physics beyond the SM can significantly enhance
this parameter~\cite{fermilab}.

In this context, experiments have made significant progress to better
understand contributions to the decay $B^0_s\to J/\psi\,K^+K^-$
beyond $B^0_s\to J/\psi\,\phi(1020)(\to K^+K^-)$.
A recent discovery in this field is the decay $B^0_s\to J/\psi
f'_2(1525)$, whose branching fraction relative to $B^0_s\to
J/\psi\,\phi(1020)$ is measured to be $(26.4\pm 2.7\left(\mathrm{stat}\right)\pm 2.4\left(\mathrm{syst}\right))\%$ by
LHCb~\cite{lhcb} and $(22\pm 5\left(\mathrm{stat}\right)\pm 4\left(\mathrm{syst}\right))\%$ by D\O~\cite{d0}. A first 
measurement of the entire $B^0_s\to J/\psi\,K^+K^-$ decay rate 
(including resonant and non-resonant decays) was recently performed by LHCb with a 
measured branching fraction of $(7.70\pm0.08\left(\mathrm{stat}\right)\pm0.39\left(\mathrm{syst}\right)\pm0.60
\left(f_s/f_d\right))\times 10^{-4}$ \cite{lhcb_new}.

In this analysis, we study the decay $B^0_s\to J/\psi\,K^+K^-$ using
the Belle data 
and determine its
absolute branching fraction. We identify the resonant contributions
$B^0_s\to J/\psi\,\phi(1020)(\to K^+K^-)$ and $B^0_s\to J/\psi\,f'_2(1525)(\to
K^+K^-)$ and determine the $S$-wave contribution in the
$\phi(1020)$~mass region. In contrast to hadron collider experiments, we normalize to the 
absolute number of $B^0_s\bar{B}^0_s$ pairs produced rather than to a reference decay channel.
In addition, to determine the $S$-wave contribution in the $\phi(1020)$
     mass region, we fit to the $K^{+}K^{-}$ mass distribution rather
     than perform an angular analysis. Thus, our results are
     obtained using methods with systematic
         uncertainties that both differ from previous analyses.

\section{Experimental Procedure}

\subsection{Data Sample and Event Selection}

The data used in this analysis were taken with the Belle
detector~\cite{Belle} at the KEKB asymmetric-energy
$e^+e^-$~collider~\cite{KEKB}. Belle is a large-solid-angle magnetic
spectrometer that consists of a silicon vertex detector (SVD),
a 50-layer central drift chamber (CDC), an array of
aerogel threshold Cherenkov counters (ACC),  
a barrel-like arrangement of time-of-flight
scintillation counters (TOF), and an electromagnetic calorimeter
comprised of CsI(Tl) crystals (ECL) located inside 
a superconducting solenoid coil that provides a 1.5~T
magnetic field.  An iron flux-return located outside of
the coil is instrumented to detect $K_L^0$ mesons and to identify
muons (KLM).

The Belle data sample taken at the $\Upsilon(5S)$~resonance has
an integrated luminosity of 121.4~fb$^{-1}$ and contains $\left(7.1\pm1.3\right)$ million
$B^0_s\bar B^0_s$~events with a cross section for the process
  $e^+e^-\to b\bar b$ of $\sigma_{b\bar b}=(0.340\pm 0.016)$~nb and a
  fraction of $b\bar b$~states hadronizing into $B^{(*)}_s\bar
  B^{(*)}_s$ of $f_s = (17.2\pm 3.0)\%$~\cite{Esen}.

Monte Carlo (MC) simulated events equivalent
to at least six times the integrated luminosity of the data are used to evaluate the signal acceptance 
and perform background studies. MC events are generated with EvtGen ~\cite{Lange:2001uf}, and a 
full detector simulation based on GEANT3 ~\cite{Brun:1987ma} is
applied. QED bremsstrahlung is included using the PHOTOS
package~\cite{Barberio:1993qi}. 

Hadronic events are selected based on the
charged track multiplicity and the visible energy in the calorimeter. Charged tracks are required to originate from within 4 ~cm along the beam axis and 0.5 ~cm in the transverse plane with respect to the $e^+e^-$~interaction point.  Electron candidates are identified using the
ratio of the energy detected in the ECL to the track momentum, the ECL
shower shape, position matching between track and ECL cluster, the
energy loss in the CDC ($dE/dx$), and the response of
the ACC~counters. Muons are identified based on their penetration
range and transverse scattering in the KLM~detector. Kaon candidates
are distinguished from pion tracks by using combined information from
the CDC, the ACC and the TOF scintillation counters.

To reconstruct $J/\psi$~mesons, two identified leptons with the same flavor ($e$ or $\mu$)
and opposite charges are combined. The energy loss from bremsstrahlung is
partially recovered by adding back the four-momentum of any photon within a 5$^\circ$~cone
around the electron or positron direction. The invariant masses of the
$J/\psi\to e^+e^-(\gamma)$ and $J/\psi\to\mu^+\mu^-$ candidates are
required to lie within the range $2.946~\mathrm{GeV} <
M(e^+e^-(\gamma)) < 3.133~\mathrm{GeV}$ and
$3.036~\mathrm{GeV} < M(\mu^+\mu^-)<
3.133~\mathrm{GeV}$, respectively. The $J/\psi$ mass resolution is
approximately 11~MeV in the
electron channel and about 10~MeV in the muon channel. Asymmetric mass windows are used to accommodate
the residual bremsstrahlung tails.

The $J/\psi$~candidate is combined with two oppositely charged kaon candidates to form a $B^0_s$~candidate. We accept candidates in the entire $K^+K^-$ phase space. 
The resolution in $M(K^+K^-)$ is approximately 1~MeV.
Due to
the two-body kinematics in the process $e^+e^-\to\Upsilon(5S)\to
B^{(*)}_s\bar B^{(*)}_s$, the $B^0_s$ signal is extracted using
the following two kinematic variables: the energy difference $\Delta
E=E^*_B-E^*_\mathrm{beam}$ and the beam-energy constrained mass
$M_\mathrm{bc}=\sqrt{E^{*2}_\mathrm{beam}-(\vec p^{\;*}_B)^2}$, where
$E^*_\mathrm{beam}$ is the beam energy in the center-of-mass (c.m.)
frame of the colliding beams, and $E^*_B$ and $\vec p^{\;*}_B$ denote the
energy and the momentum of the reconstructed $B^0_s$~meson,
respectively, in the c.m.\ system.
As the photon from the decay $B^*_s\to B^0_s\gamma$ is not
reconstructed, there are three signal regions in the
$(M_\mathrm{bc},\Delta E)$~plane, corresponding to the three initial
states $B^0_s\bar B^0_s$, $B^0_s\bar B^*_s$ (or $B^*_s\bar B^0_s$) and
$B^*_s\bar B^*_s$. We select the most abundant initial state
$B^*_s\bar B^*_s$ (the $B^*_s\bar B^*_s$ fraction in $B^{(*)}_s\bar
B^{(*)}_s$ events being $f_{B^*_s\bar B^*_s} = \left(87.0\pm
1.7\right)\%$~\cite{Esen}) by requiring $-0.2~\mathrm{GeV}<\Delta
E< 0.1~\mathrm{GeV}$ and $M_\mathrm{bc}> 5.4~\mathrm{GeV}$, as
  the signal peaks around $\Delta E = M(B^*_s) - M(B^0_s)
  \approx 0.049~\mathrm{GeV}$ and $M_\mathrm{bc} = M(B^*_s)\approx
  5.415~\mathrm{GeV}$ for the $B^*_s\bar B^*_s$ signal region.

\subsection{Backgrounds and Signal Extraction}

Background to the $B^0_s\to J/\psi\,K^+K^-$~signal arises from
random combinations in $\Upsilon(5S)$~events and from so-called
continuum, {\it i.e.}, events originating from the process $e^+e^-\to
q\bar q$ with $q=u,d,s$ or $c$. Contributions from the latter are
suppressed by exploiting the difference in event shape between
$\Upsilon(5S)$ and continuum events (spherical vs.\ jet-like,
respectively) and requiring the ratio of the second to zeroth
Fox-Wolfram moment
$R_2 = H_2 / H_0$~\cite{FW} to be less than 0.4. This selection was
optimized for this decay topology in the analysis of the decay
$B^0_s\to J/\psi\,\pi^+\pi^-$~\cite{lifin_f0}.

Signal extraction is performed independently for the $J/\psi\to e^+e^-$ and 
$J/\psi\to\mu^+\mu^-$ subsamples by a two-dimensional unbinned maximum
likelihood fit in $\Delta E$ and $M(K^+K^-)$. The fit range is $-0.2~\mathrm{GeV}<\Delta
E< 0.1~\mathrm{GeV}$ and $0.95~\mathrm{GeV} < M(K^+K^-) < 2.4~\mathrm{GeV}$ and 
takes into account resolution effects at the lower end of the $M(K^+K^-)$ phase space. The probability density
function (PDF) for signal~\cite{signal} in $\Delta E$ is
parameterized with a sum of a Gaussian and a Crystal Ball~\cite{CB}
function (a sum of two Gaussian functions) for the $J/\psi\to
e^+e^-$
($J/\psi\to\mu^+\mu^-$) data sample. The parameters of these PDFs\ are
determined from data using a control sample of $B^0\to J/\psi
K^{*}(892)^0$ decays with $K^{*}(892)^0\to K^+\pi^-$. The signal shapes of the
$\phi(1020)$ and the $f'_2(1525)$~resonances in $M(K^+K^-)$ are each described by a
non-relativistic Breit-Wigner function whose width includes both
  the natural width and the detector resolution. The remaining
$J/\psi(K^+K^-)_\mathrm{other}$ component is modeled with an ARGUS
function~\cite{argus} in $M(K^+K^-)$. When we perform the fit on the data, the shape 
parameters of all signal PDFs for the $\Delta E$ distribution are fixed using the control sample, while 
the parameters of the signal PDFs for the $M(K^+K^-)$ distribution are fixed using MC simulations.

The background, which includes contributions from
combinatorial background in $\Upsilon(5S)$~events and continuum
background, is parameterized by a first-order polynomial in $\Delta E$
and an ARGUS function in $M(K^+K^-)$. The parameters of the background
PDFs\ are determined from a data sideband defined by
$5.25~\mathrm{GeV}< M_\mathrm{bc}< 5.35~\mathrm{GeV}$ and fixed in the fit on the real data.

The entire signal extraction procedure has been tested and validated
on simulated events. Terms in the PDF due to interference among the 
$J/\psi\,\phi(1020)$,  $J/\psi\,f'_2(1525)$ and $J/\psi(K^+K^-)_\mathrm{other}$ components 
cancel after integration over angular variables, since the $K^+K^-$ systems have distinct quantum numbers of $1$, $2$ and $0$, respectively. As we find that the angular acceptance is 
approximately flat within our statistics, we do not consider interference effects among these components.

\begin{figure}[htb]
  \includegraphics[width=1.0\columnwidth]{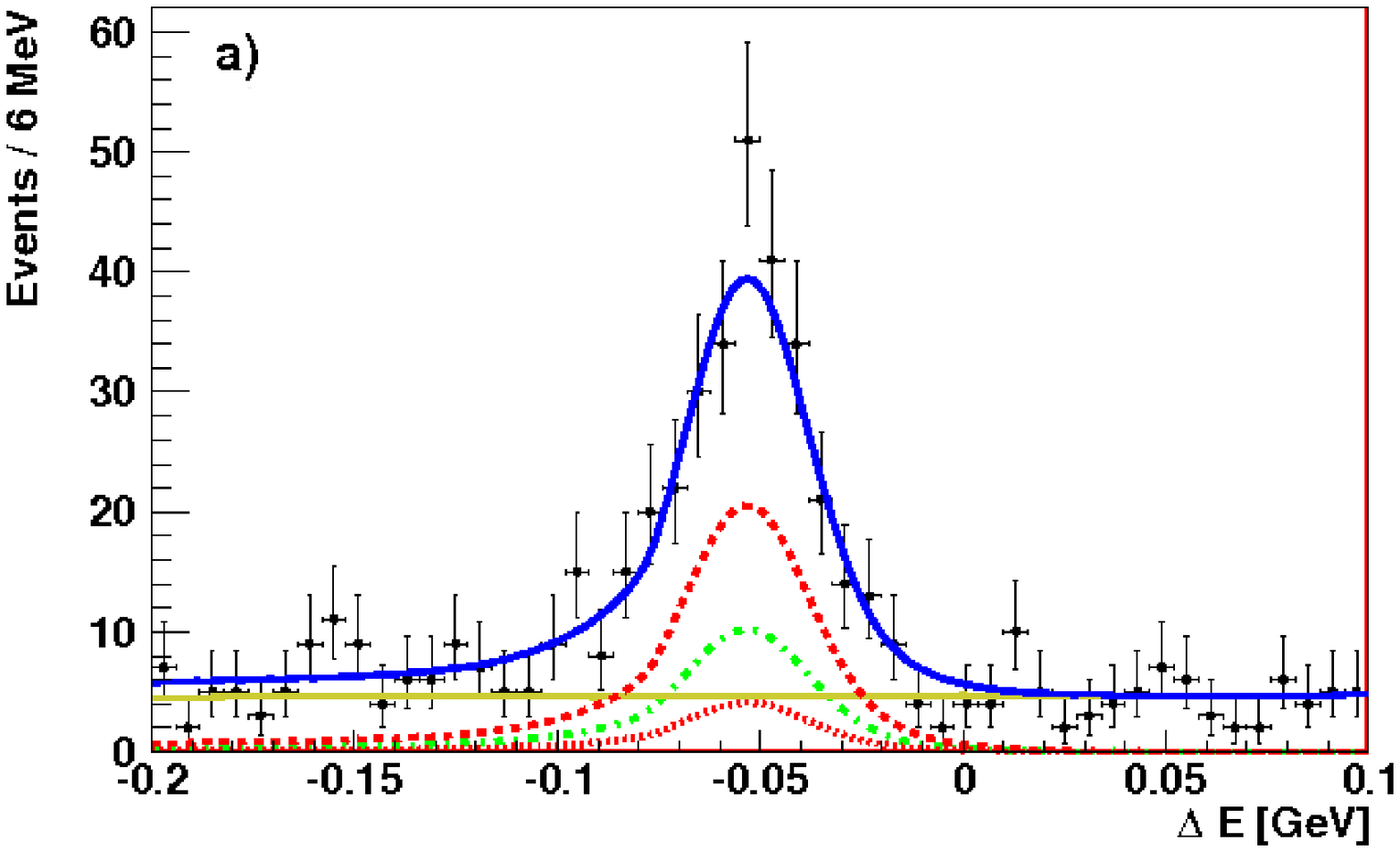}\\
  \includegraphics[width=1.0\columnwidth]{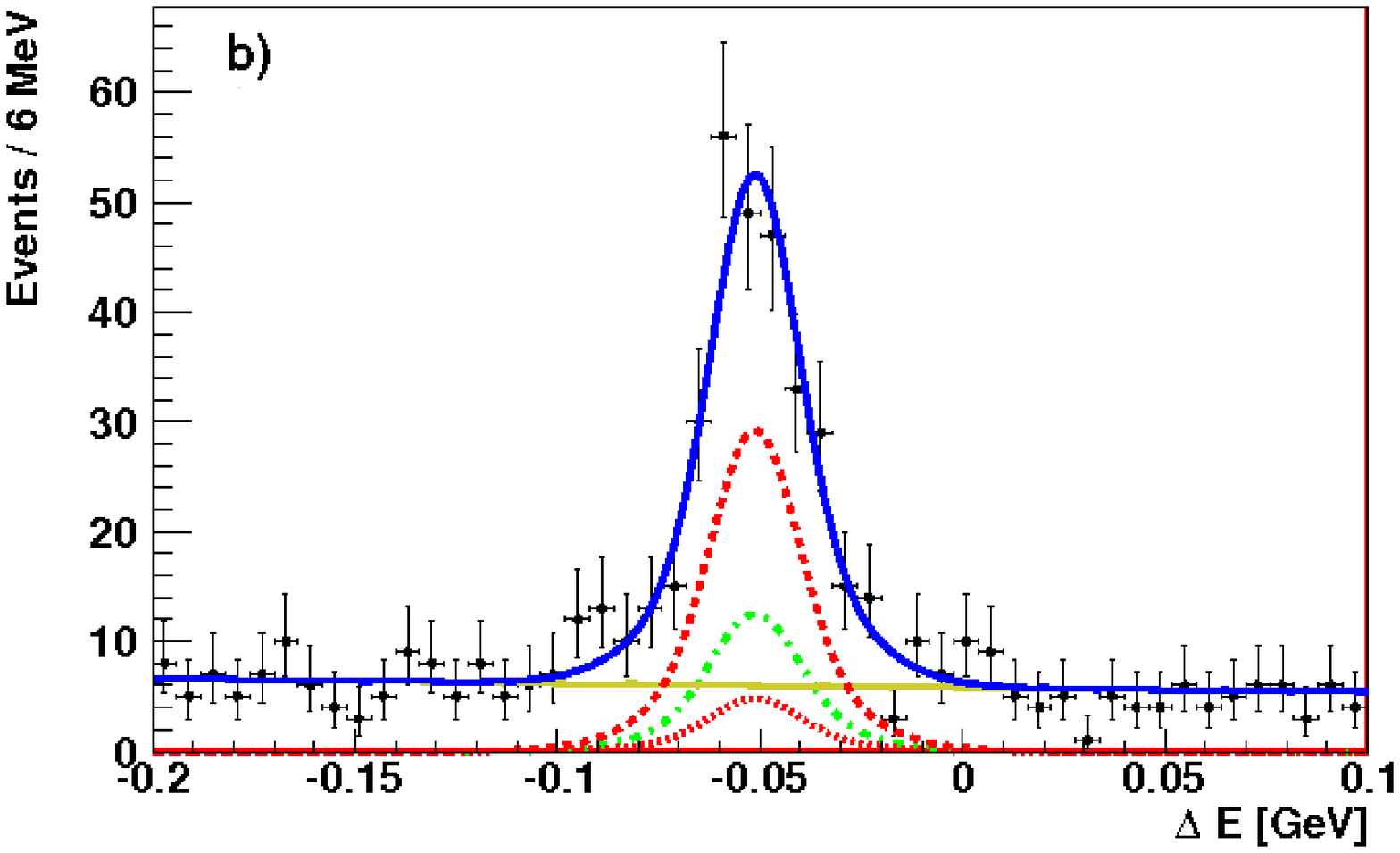}
  \caption{Projection of the fit in $\Delta E$ for a) $J/\psi\to
    e^+e^-$ and b) $J/\psi\to\mu^+\mu^-$ events. The black points with error bars are the data and the upper solid line corresponds to the entire PDF\ model. From top to bottom, the
    peaking components are $J/\psi\,\phi(1020)$, $J/\psi
    (K^+K^-)_\mathrm{other}$ and $J/\psi\,f'_2(1525)$. Background is
    shown by the lower solid line.} \label{fig:de}
\end{figure}

\begin{figure}[!]
  \includegraphics[width=1.0\columnwidth]{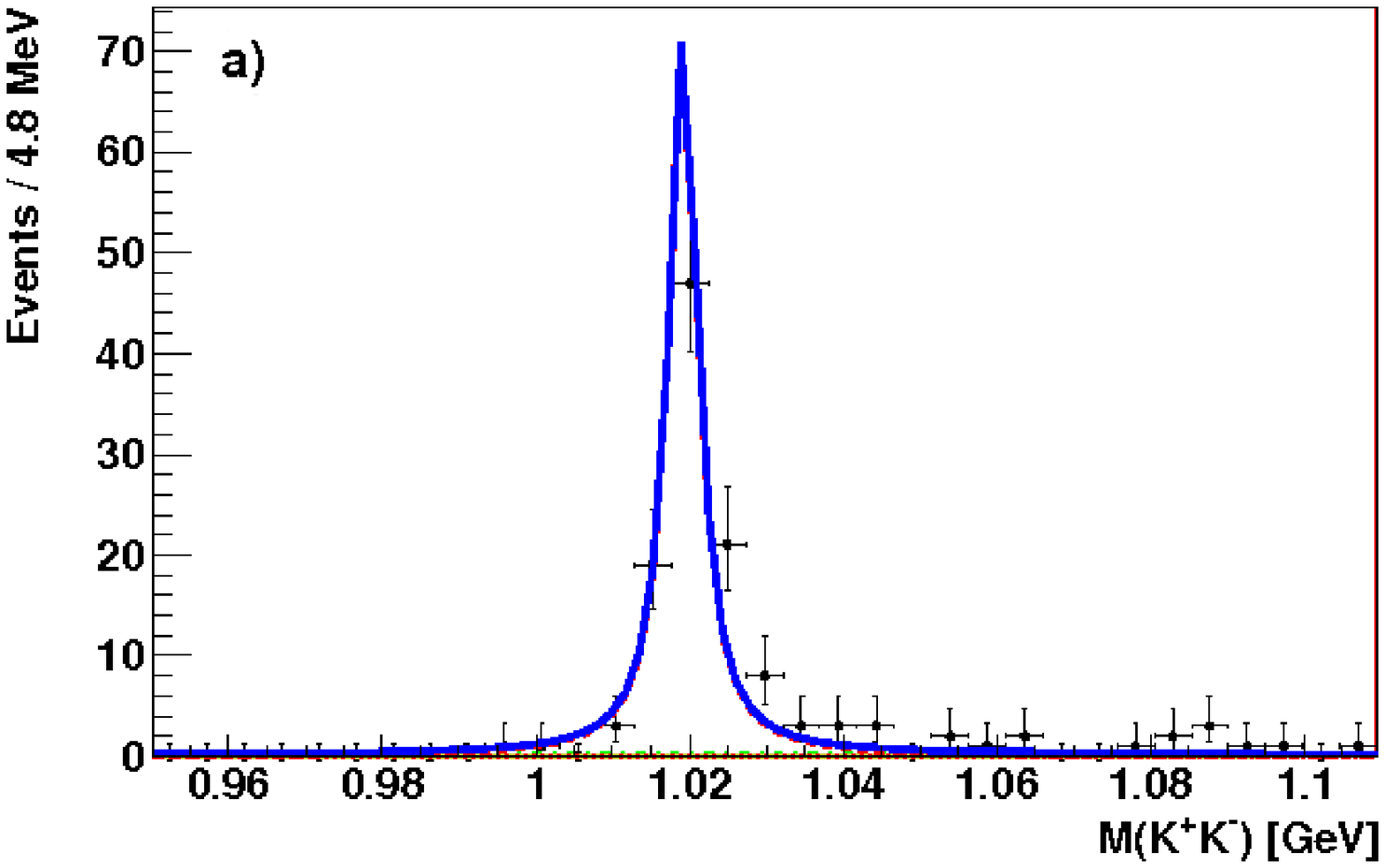}\\
  \includegraphics[width=1.0\columnwidth]{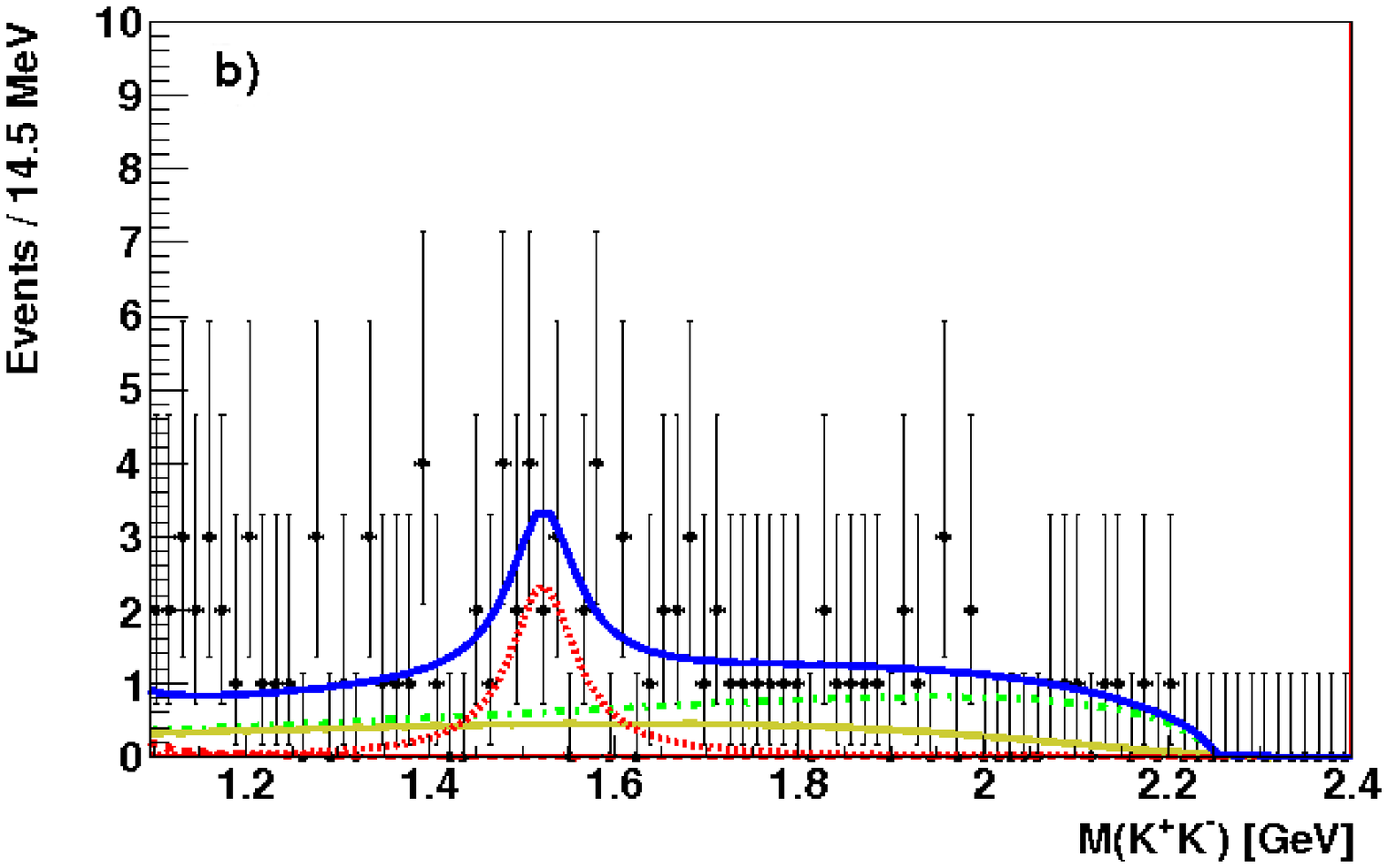}\\
  \includegraphics[width=1.0\columnwidth]{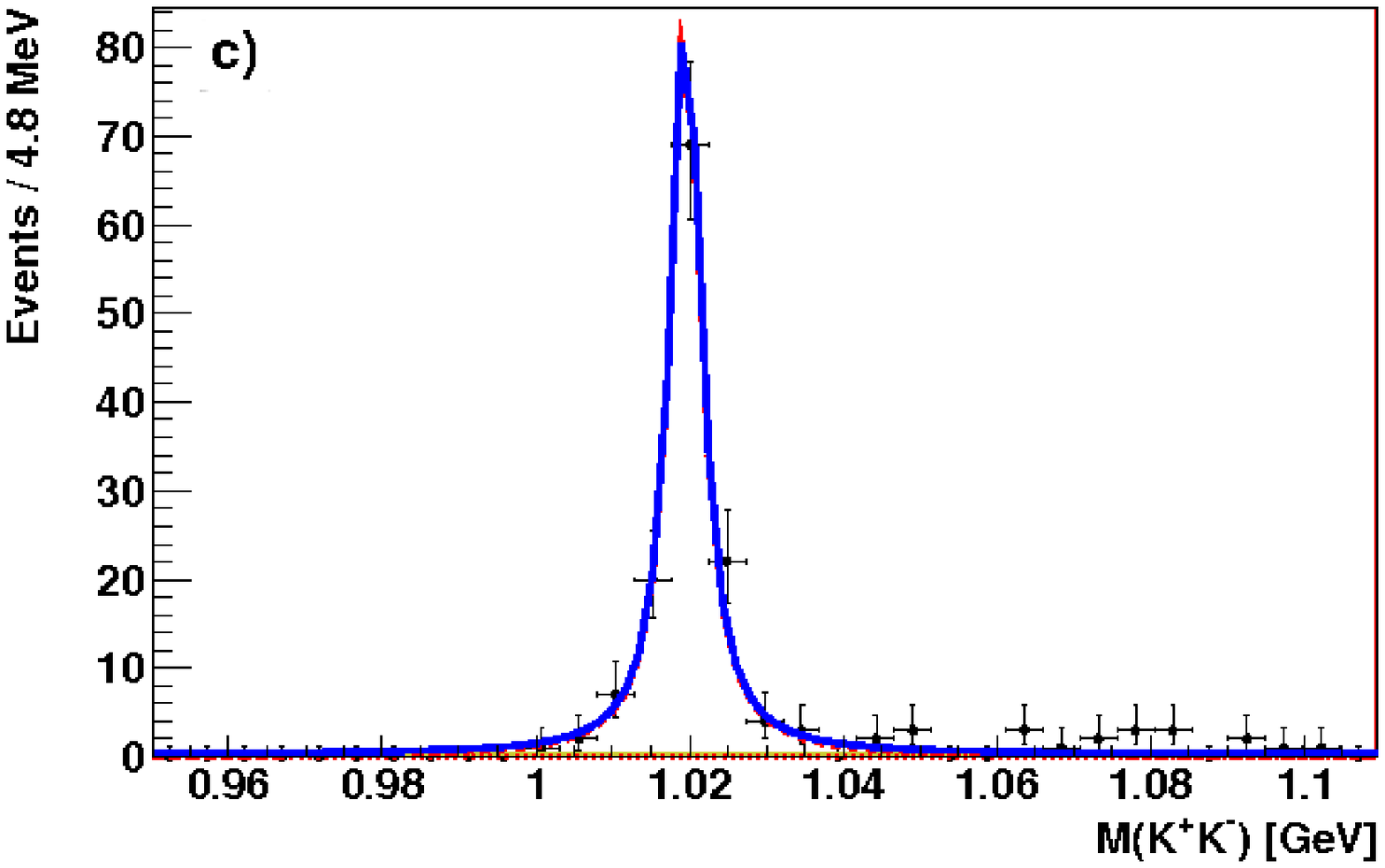}\\
  \includegraphics[width=1.0\columnwidth]{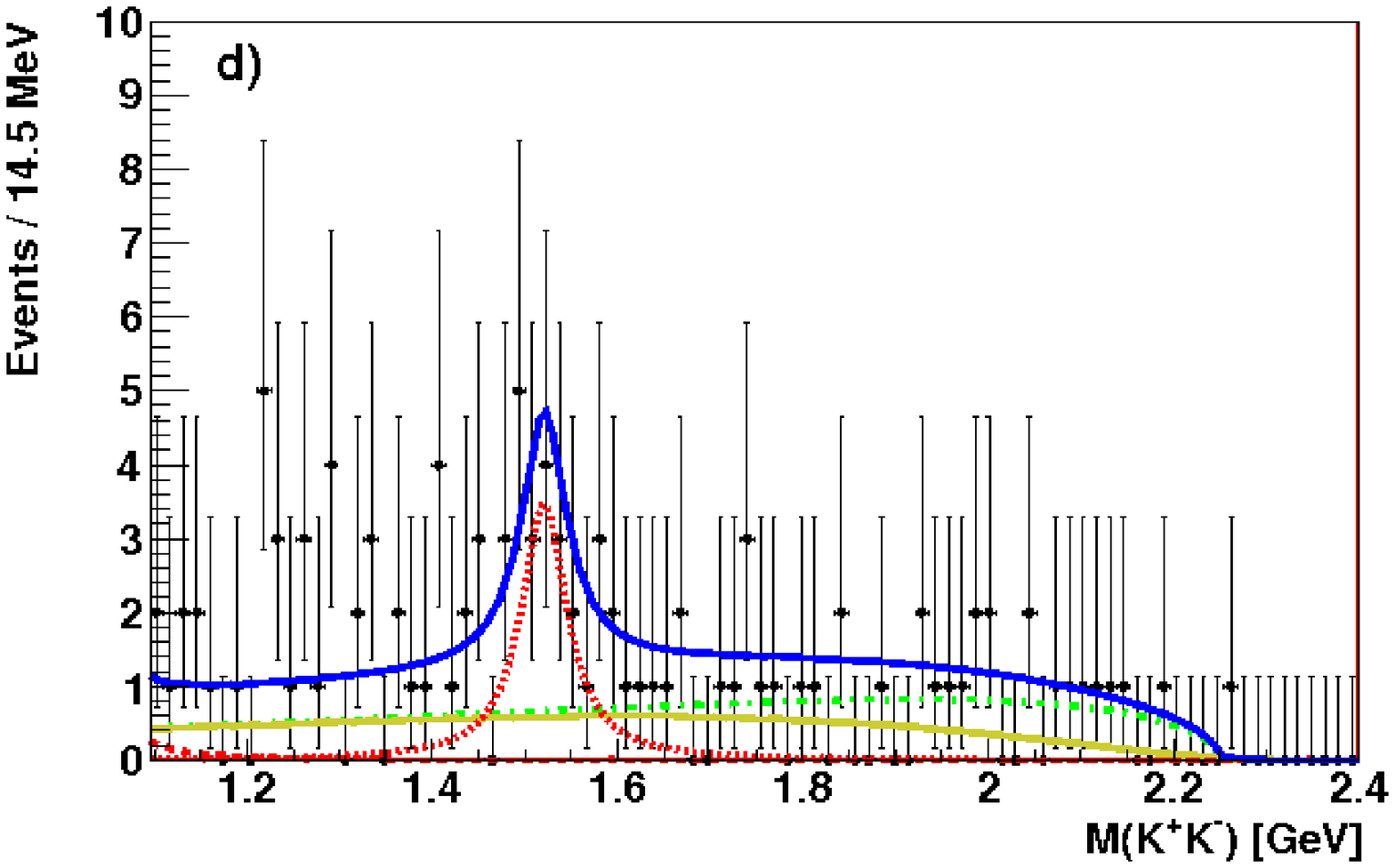}
  \caption{Projection of the fit in $M(K^+K^-)$ for events in the signal range
    $-0.07~\mathrm{GeV}<\Delta E< -0.03~\mathrm{GeV}$. Panels a) and b) show 
the $\phi(1020)$ and $f'_2(1525)$~mass regions, respectively, for $J/\psi\to e^+e^-$~events; 
panels c) and d) are the same for $J/\psi\to\mu^+\mu^-$~events. In all
    plots, the upper solid line corresponds to the entire
    PDF\ model, which overlaps with the curve of the $J/\psi\,\phi(1020)$ component in a) and c). The signal components and the background are
    shown using the line style of Fig. \ref{fig:de}.} \label{fig:mkk}
\end{figure}

The yields
obtained for the $J/\psi\to e^+e^-$ and $J/\psi\to\mu^+\mu^-$ samples
are given in Table~\ref{tab:fit_results}. Figures~\ref{fig:de} and
\ref{fig:mkk} show the projections of the fit in $\Delta E$
and $M(K^+K^-)$, respectively.
\begin{table}[htb]
  \caption{Extracted yields for signal components and background in the
    $J/\psi\to e^+e^-$ and $J/\psi\to\mu^+\mu^-$
    samples.} \label{tab:fit_results}
  \begin{tabular}
    {@{\hspace{0.5cm}}l@{\hspace{0.5cm}}
      @{\hspace{0.5cm}}c@{\hspace{0.5cm}}
      @{\hspace{0.5cm}}c@{\hspace{0.5cm}} }
    \hline \hline
    Channel & $e^+e^-$ & $\mu^+\mu^-$ \\
    \hline
    $J/\psi\,\phi(1020)$ & $168\pm 13.5$ & $158\pm 13$ \\
    $J/\psi\,f'_2(1525)$ & $34\pm 10$ & $26\pm 8$ \\
    $J/\psi(K^+K^-)_\mathrm{other}$ & $83\pm 17$ & $67\pm 14$ \\
    Background & $232\pm 19$ & $300\pm 20$ \\
    \hline \hline
  \end{tabular}
\end{table}
\begin{table}[htb]
  \caption{Reconstruction efficiency $\epsilon$ for all three
    investigated decay modes. The quoted error corresponds to the
      uncertainty due to the MC statistics.} 
\label{tab:efficiency}
  \begin{tabular}
    {@{\hspace{0.5cm}}l@{\hspace{0.5cm}}
      @{\hspace{0.5cm}}c@{\hspace{0.5cm}}
      @{\hspace{0.5cm}}c@{\hspace{0.5cm}} }
    \hline \hline
    Channel & $\epsilon_{e^+e^-}\:[\%]$ & $\epsilon_{\mu^+\mu^-}\:[\%]$ \\
    \hline
    $J/\psi\,\phi(1020)$ & $31.0\pm 0.1$ & $33.2\pm 0.1$ \\
    $J/\psi\,f'_2(1525)$ & $28.4\pm 0.2$ & $30.5\pm 0.2$ \\
    $J/\psi(K^+K^-)_\mathrm{other}$ & $29.7\pm 0.1$ & $32.5\pm 0.1$ \\
    \hline \hline
  \end{tabular}
\end{table}

\section{Results and Systematic Uncertainties}

The absolute branching fraction for the decay $B^0_s\to
J/\psi\,\phi(1020)$ is calculated from the fitted yields in
Table~\ref{tab:fit_results} as
\begin{multline}
  \mathcal{B}[B^0_s\to
  J/\psi\,\phi(1020)]=\\
  \frac{N_{J/\psi\,\phi(1020)}}{2\mathcal{L}\sigma_{b\bar
      b}f_s f_{B^*_s\bar B^*_s}\epsilon\,\mathcal{B}[J/\psi\to\ell^+\ell^-]\mathcal{B}[\phi(1020)\to
    K^+K^-]}~, \label{eq:1}
\end{multline}
where $N_{J/\psi\,\phi(1020)}$ is the extracted yield, $\mathcal{L}$ is the
luminosity of the Belle $\Upsilon(5S)$~sample,
and
$\mathcal{B}[J/\psi\to\ell^+\ell^-]$ and $\mathcal{B}[\phi(1020)\to K^+K^-]$
are the sub-decay branching fractions. The parameter $\epsilon$
  denotes the reconstruction efficiency, whose values are given in Table \ref{tab:efficiency}. Applying this formula to the
electron and muon samples and averaging the results, we obtain

\begin{multline}
  \mathcal{B}[B^0_s\to J/\psi\,\phi(1020)]=\\
  (1.25\pm 0.07\left(\mathrm{stat}\right)\pm
  0.08\left(\mathrm{syst}\right)\pm 0.22\left(f_s\right))\times 10^{-3}~, \label{eq:2}
\end{multline}
where the first uncertainty is statistical, the second is systematic and
the third is due to the uncertainty in $f_s$. Similarly,
we obtain for $B^0_s\to J/\psi\,f'_2(1525)$
\begin{multline}
  \mathcal{B}[B^0_s\to J/\psi\,f'_2(1525)]=\\
  (0.26\pm
  0.06\left(\mathrm{stat}\right)\pm 0.02\left(\mathrm{syst}\right)\pm 0.05\left(f_s\right))\times
  10^{-3}~. \label{eq:3}
\end{multline}
The branching fraction ratio is
\begin{equation}
  \frac{\mathcal{B}[B^0_s\to J/\psi\,f'_2(1525)]}{\mathcal{B}[B^0_s\to
    J/\psi\,\phi(1020)]}=(21.5\pm 4.9\left(\mathrm{stat}\right)
\pm2.6\left(\mathrm{syst}\right))\%~. \label{eq:4}
\end{equation}
The significance of the $B^0_s\to J/\psi\,f'_2(1525)$ signal is
equal to 3.3 standard deviations (including the systematic
uncertainty), which is calculated from the difference of the
  log-likelihood values of the default fit and a fit including
  background components only.
The branching fraction for the entire $B^0_s\to J/\psi
K^+K^-$~component (including the non-resonant decay and the resonant contributions 
$B^0_s\to J/\psi\,\phi(1020)$ and $B^0_s\to J/\psi\,f'_2(1525)$) is
\begin{multline}
  \mathcal{B}[B^0_s\to J/\psi\,K^+K^-]=\\
  (1.01\pm 0.09 \left(\mathrm{stat}\right)\pm
  0.10 \left(\mathrm{syst}\right)\pm 0.18\left(f_s\right))\times 10^{-3}. 
\label{eq:5}
\end{multline}

The contributions to the systematic uncertainties given in
  Eqs.~\ref{eq:1}$-$\ref{eq:5} are
listed in Table~\ref{tab:sys} and fall into three categories:
uncertainties in the input parameters in Eq.~\ref{eq:1}, uncertainties
related to signal and detector response simulation, and the 
PDF\ model. The first class of uncertainties is dominant because of
the large uncertainty in $f_s$, which we quote separately. 

To estimate the error related to the $\phi(1020)$~polarization in the
simulation of $B^0_s\to J/\psi\,\phi(1020)$, we use the difference in
efficiency between a simulation using the polarization parameters
determined by CDF~\cite{cdfswave} and a different sample giving equal 
weights to each helicity amplitude.

The systematic error related to lepton ($\ell^{\pm}$) identification is
  determined using $\gamma\gamma\to\ell^{+}\ell^{-}$ events. The 
uncertainty arising from kaon identification is
     determined from a sample of $D^{*+}\to D^{0} \pi^{+}$, $D^{0}\to K^{-}\pi^{+}$ decays.

The error related to the PDF\ parameters is obtained by
performing 1000 pseudo-experiments, sampling each
     parameter from a Gaussian distribution having a mean value and
     width equal to the parameter's central value and uncertainty.
The width of the distribution of signal yields is taken as the systematic uncertainty. As we do not find significant impact from the correlations among the parameters, we sample each parameter independently when performing this calculation.

The systematic error due to the PDF\ model is estimated by repeating the fit with alternative
PDF\ functions, including a relativistic Breit-Wigner function and a
non-relativistic Breit-Wigner function with a phase space correction for
the $\phi(1020)$ and $f'_2(1525)$ resonances, and a pure phase space
description for the $J/\psi(K^+K^-)_\mathrm{other}$ component. The
maximum deviation between these fit results and the results obtained with
the default model is taken as a systematic uncertainty. Only for the $J/\psi(K^+K^-)_\mathrm{other}$ component was a significant deviation found. For the $J/\psi\,\phi(1020)$ and the 
$J/\psi\,f'_2(1525)$ components the deviations were negligibly small.

\begin{table}[htb]
  \caption{Contribution to the systematic uncertainty in the $B^0_s\to
    J/\psi\,K^+K^-$ branching fractions.} \label{tab:sys}	
  \begin{tabular}
    {l c} 
    \hline \hline
    Source & Uncertainty $\left[\%\right]$ \\
    \hline
    Luminosity  & 0.7\\
    $\sigma_{b\bar b}$~\cite{Esen} & 4.7\\
    $f_s$~\cite{Esen} & 17.4\\
    $f_{B^*_s\bar B^*_s}$ ($B^*_s\bar B^*_s$ fraction in $B^{(*)}_s\bar B^{(*)}_s$) & 2.0\\
    $\mathcal{B}[J/\psi\to\ell^+\ell^-]$~\cite{pdg} & 1.0\\
    $\mathcal{B}[\phi(1020)\to K^+K^-]$~\cite{pdg} & 1.0\\
    $\mathcal{B}[f'_2(1525)\to K^+K^-]$~\cite{pdg} & 2.5\\
    \hline
    MC statistics & 0.3-0.8\\
    $\phi(1020)$ polarization & 1.3\\
    Charged tracking & 1.4\\
    Electron identification  & 3.1 \\
    Muon identification & 3.0 \\
    Kaon identification  & 1.9\\
    \hline
    PDF\ parameters: & \\
    $J/\psi_{e^+e^-}\phi(1020)$ & 1.1\\
    $J/\psi_{\mu^+\mu^-}\phi(1020)$ & 1.0\\
    $J/\psi_{e^+e^-}(K^+K^-)_\mathrm{other}$    & 9.1\\
    $J/\psi_{\mu^+\mu^-}(K^+K^-)_\mathrm{other}$ & 5.8\\
    $J/\psi_{e^+e^-}f'_2(1525)$ & 7.8\\
    $J/\psi_{\mu^+\mu^-}f'_2(1525)$ & 5.4\\
    PDF\ model: & \\
    $J/\psi_{e^+e^-}(K^+K^-)_\mathrm{other}$ & 2.4\\
    $J/\psi_{\mu^+\mu^-}(K^+K^-)_\mathrm{other}$ & 3.0\\
    \hline \hline
  \end{tabular}
\end{table}

The total systematic error is calculated separately for the electron and muon
channels by adding all components in quadrature. For the calculation of the weighted mean value,
the systematic errors are treated as fully correlated. 

As an additional result, the $S$-wave contribution in the $\phi(1020)$~mass
region is calculated using the signal yields presented in
Table~\ref{tab:fit_results}. Here, we assume that the $K^+K^-$ system in
$B^0_s\to J/\psi(K^+K^-)_\mathrm{other}$ is a pure $S$-wave. This
assumption is supported by the observed helicity angle distribution of
$J/\psi(K^+K^-)_\mathrm{other}$, where the helicity angle is defined 
as the angle between the $K^+$ meson and the $B^0_s$ meson in the $K^+K^-$ rest frame. Hence, the $S$-wave fraction
($S$) is the fitted yield of $J/\psi(K^+K^-)_\mathrm{other}$~events
relative to the yield of $J/\psi\,K^+K^-$ within a specific mass range,
\begin{equation}
  S = \frac{\alpha N[J/\psi(K^+K^-)_\mathrm{other}]}{\alpha
    N[J/\psi(K^+K^-)_\mathrm{other}]+\beta N[J/\psi\,\phi(1020)]}~,
\end{equation}
where $\alpha$ and $\beta$ denote the fractions of $J/\psi(K^+K^-)_\mathrm{other}$ and
       $J/\psi\,\phi(1020)$, respectively, within the
        mass range considered.
$N[J/\psi(K^+K^-)_\mathrm{other}]$ and $N[J/\psi\,\phi(1020)]$
are the fitted yields from Table~\ref{tab:fit_results}. The results are
shown in Table~\ref{tab:swave} for the mass ranges used in
hadron collider experiments. While the statistical uncertainty is
propagated via the fitted yields, the systematic uncertainty in the
$S$-wave contribution due to the PDF\ parameterization uncertainties and  the PDF model are propagated
through $\alpha$ and $\beta$.

To estimate the systematic uncertainty due to a possible 
$B^0_s\to J/\psi f_0(980)$ contribution, as seen, {\it e.g.}, by
LHCb~\cite{lhcb_new}, we investigate the difference between the PDF model used in this analysis 
and the PDF model used by LHCb, which describes the $B^0_s\to J/\psi f_0(980)$ component with 
a Flatt\'e function in $M(K^+K^-)$. From the $S$-wave contribution of $1.1\%$ obtained 
by LHCb, we calculate the value of the parameter $\alpha$ in the LHCb PDF model. We find an increase in $\alpha$ from $0.8\%$ ($1.0\%$) to $5.0\%$ for the CDF (LHCb) mass range in the electron channel, and from $0.9\%$ ($1.1\%$) to $5.0\%$ in the muon channel. 
We assign this variation as an additional model uncertainty, which we quote 
separately in Table \ref{tab:swave}.

\begin{table}[htb]
 \caption{The $J/\psi\,K^+K^-$ $S$-wave contribution in different mass
   regions around the $\phi(1020)$~resonance. The first error is statistical, the second systematic and the third error is the uncertainty due to a possible $B^0_s\to J/\psi f_0(980)$ contribution.} \label{tab:swave}
\begin{tabular}
  {l@{\hspace{0.5cm}}c}
  \hline \hline
  Mass range & 1.009~GeV $-$ 1.028~GeV\\
  CDF~\cite{cdfswave}  & $(0.8\pm 0.2)\%$\\
  This analysis & $(0.47\pm 0.07\pm 0.22 ^{+2.2}_{-0})\%$\\
  \hline
  Mass range & 1.007~GeV $-$ 1.031~GeV\\
  LHCb~\cite{lhcb_new} & $(1.1\pm 0.1 ^{+0.2}_{-0.1})\%$\\
  This analysis & $(0.57\pm 0.09\pm 0.26 ^{+2.0}_{-0})\%$\\
  \hline \hline
\end{tabular}
\end{table}

\section{Summary}

In summary, we present a  measurement of the absolute
branching fraction for the decay $B^0_s\to J/\psi\,\phi(1020)$
(Eq.~\ref{eq:2}). This result is in good agreement with the CDF Run I
result~\cite{pdg,cdf_br} as well as their preliminary results based on the
full data sample~\cite{cdf_ichep} and the current LHCb result \cite{lhcb_new}. We obtain evidence for the decay
$B^0_s\to J/\psi\,f'_2(1525)$ (Eqs.~\ref{eq:3} and \ref{eq:4}), in good
agreement with
the measurements by LHCb~\cite{lhcb,lhcb_new} and D\O~\cite{d0}. We also present a measurement of the
entire $B^0_s\to J/\psi\,K^+K^-$ component including resonant and
non-resonant decays (Eq.~\ref{eq:5}). Finally, we determine the $S$-wave 
fraction of $B^0_s\to J/\psi\,K^+K^-$ in the $\phi(1020)$~mass region
(Table~\ref{tab:swave}). Our central value is somewhat lower than the LHCb and CDF values but in agreement with their results when including the systematic error due to a possible 
$B^0_s\to J/\psi f_0(980)$ component.

\section*{Acknowledgments}

We thank the KEKB group for the excellent operation of the
accelerator; the KEK cryogenics group for the efficient
operation of the solenoid; and the KEK computer group,
the National Institute of Informatics, and the 
PNNL/EMSL computing group for valuable computing
and SINET4 network support.  We acknowledge support from
the Ministry of Education, Culture, Sports, Science, and
Technology (MEXT) of Japan, the Japan Society for the 
Promotion of Science (JSPS), and the Tau-Lepton Physics 
Research Center of Nagoya University; 
the Australian Research Council and the Australian 
Department of Industry, Innovation, Science and Research;
Austrian Science Fund under Grant No. P 22742-N16;
the National Natural Science Foundation of China under
contract No.~10575109, 10775142, 10875115 and 10825524; 
the Ministry of Education, Youth and Sports of the Czech 
Republic under contract No.~MSM0021620859;
the Carl Zeiss Foundation, the Deutsche Forschungsgemeinschaft
and the VolkswagenStiftung;
the Department of Science and Technology of India; 
the Istituto Nazionale di Fisica Nucleare of Italy; 
The BK21 and WCU program of the Ministry Education Science and
Technology, National Research Foundation of Korea Grant No.\ 
2010-0021174, 2011-0029457, 2012-0008143, 2012R1A1A2008330,
BRL program under NRF Grant No. KRF-2011-0020333,
and GSDC of the Korea Institute of Science and Technology Information;
the Polish Ministry of Science and Higher Education and 
the National Science Center;
the Ministry of Education and Science of the Russian
Federation and the Russian Federal Agency for Atomic Energy;
the Slovenian Research Agency;
the Basque Foundation for Science (IKERBASQUE) and the UPV/EHU under 
program UFI 11/55;
the Swiss National Science Foundation; the National Science Council
and the Ministry of Education of Taiwan; and the U.S.\
Department of Energy and the National Science Foundation.
This work is supported by a Grant-in-Aid from MEXT for 
Science Research in a Priority Area (``New Development of 
Flavor Physics''), and from JSPS for Creative Scientific 
Research (``Evolution of Tau-lepton Physics'').


\begin{thebibliography}{99}

\bibitem{Dunietz:2000cr} 
  I.~Dunietz, R.~Fleischer and U.~Nierste,

  Phys.\ Rev.\ D {\bf 63}, 114015 (2001)
  [hep-ph/0012219].

\bibitem{KM}
  M.~Kobayashi and T.~Maskawa, Prog. Theor. Phys. {\bf 49}, 652 (1973).

\bibitem{Cab}
  N. Cabibbo, Phys. Rev. Lett. {\bf 10}, 531 (1963).

\bibitem{CC}
  Throughout this paper, the inclusion of the charge-conjugate decay mode is implied.

\bibitem{cdfswave} 
T.~Aaltonen {\it et al.}  [CDF Collaboration],
 Phys.\ Rev.\ Lett. {\bf 109}, 171802 (2012)
 [arXiv:1208.2967 [hep-ex]].

\bibitem{Abazov:2011ry} 
  V.~M.~Abazov {\it et al.}  [D\O Collaboration],
  Phys.\ Rev.\ D {\bf 85}, 032006 (2012)
  [arXiv:1109.3166 [hep-ex]].

\bibitem{LHCb:2011aa} 
  R.~Aaij {\it et al.}  [LHCb Collaboration],
  Phys.\ Rev.\ D\  {\bf 87}, 112010 (2013)
  [arXiv:1304.2600 [hep-ex]].

\bibitem{atlas_phi}
G. Aad {\it et al.}  [ATLAS Collaboration], 
JHEP {\bf 1212}, 072 (2012) 
 [arXiv:1208.0572 [hep-ex]].

\bibitem{fermilab}
  K.~Anikeev {\it et al.},
  \newblock FERMILAB Report
  \newblock No. 01-197 (2001),
  \newblock and references therein.

\bibitem{lhcb}
  R.~Aaij {\it et al.}  [LHCb Collaboration],
  Phys.\ Rev.\ Lett.\  {\bf 108}, 151801 (2012)
  [arXiv:1112.4695 [hep-ex]].

\bibitem{d0} 
  V.~M.~Abazov {\it et al.}  [D\O Collaboration],
 Phys.\ Rev.\ D {\bf 86}, 092011 (2012)
  [arXiv:1204.5723 [hep-ex]].

\bibitem{lhcb_new}
  R.~Aaij {\it et al.}  [LHCb Collaboration],
  Phys.\ Rev.\ D {\bf 87}, 072004 (2013)
  [arXiv:1302.1213 [hep-ex]].

\bibitem{Belle}
  A.~Abashian {\it et al.} [Belle Collaboration], Nucl. Instr. and Meth. A
  {\bf 479}, 117 (2002);\\ also see detector section in
 J. Brodzicka et al., Prog. Theor. Exp. Phys. (2012) 04D001.

\bibitem{KEKB}
  S.~Kurokawa and E.~Kikutani, Nucl. Instr. and Meth. A  {\bf 499}, 1
  (2003), and other papers included in this volume.\\
  T. Abe {\it et al.}, Prog. Theor. Exp. Phys. (2013) 03A001 and following
 articles up to 03A011.

\bibitem{Esen}
S.~Esen {\it et al.}  [Belle Collaboration],
 Phys. Rev. D {\bf 87}, 031101 (2013)
  [arXiv:1208.0323 [hep-ex]].

%

\bibitem{Lange:2001uf}
  D.~J.~Lange,
  %
  Nucl.\ Instrum.\ Meth.\ A {\bf 462}, 152 (2001).

\bibitem{Brun:1987ma}
  R.~Brun, F.~Bruyant, M.~Maire, A.~C.~McPherson and P.~Zanarini,
  %
  CERN-DD/EE/84-1.

\bibitem{Barberio:1993qi}
  E.~Barberio and Z.~Was,
  Comput.\ Phys.\ Commun.\  {\bf 79}, 291 (1994).

\bibitem{FW}
  G.C.\ Fox and S.\ Wolfram, Phys.\ Rev.\ Lett.\ {\bf 41}, 1581
  (1978).

\bibitem{lifin_f0}
  J. Li {\it et al.} [Belle Collaboration], Phys. Rev. Lett. {\bf 106}, 121802 (2011).

\bibitem{signal}
  In this paper, ``signal'' refers to all of $B^0_s\to J/\psi\,\phi(1020)(\to K^+K^-)$,
  $B^0_s\to J/\psi\,f'_2(1525)(\to K^+K^-)$ and $B^0_s\to
  J/\psi(K^+K^-)_\mathrm{other}$. The notation $J/\psi(K^+K^-)_\mathrm{other}$ includes the non-resonant decay
  $B^0_s\to J/\psi\,K^+K^-$ as well as decays via resonant intermediate
  states, except the $\phi(1020)$ and the $f'_2(1525)$~resonance.

\bibitem{CB}
  T. Skwarnicki, Ph.D.\ Thesis, Institute for Nuclear Physics, Krakow 1986; 
  DESY Internal Report, DESY F31-86-02 (1986).

\bibitem{argus}
  H.~Albrecht {\it et al.} [ARGUS Collaboration], Phys. Lett. B{\bf 241}, 278 (1990).

\bibitem{pdg}
  J.~Beringer {\it et al.} [Particle Data Group], Phys.\ Rev.\ D {\bf
    86}, 010001 (2012) and 2013 partial update for the 2014 version.




\bibitem{cdf_br}
  F.~Abe {\it et al.}  [CDF Collaboration],
  Phys.\ Rev.\ D {\bf 54}, 6596 (1996)
  [hep-ex/9607003].

\bibitem{cdf_ichep}
  CDF~Collaboration,
  \newblock Public Note \begin{bf}10795\end{bf}, (2012).


\end{thebibliography}
\end{document}